\documentclass[11pt]{article}
\usepackage{moriond,epsfig}

\def\Journal#1#2#3#4{{#1} {\bf #2}, #3 (#4)}

\def\APJ{\em Astrophys. Journal}
\def\APJL{\em Astrophys. Journal Lett.}

\def\MNRAS{{\em MNRAS}}

\def\be{\begin{equation}}
\def\ee{\end{equation}}
\def\bea{\begin{eqnarray}}
\def\eea{\end{eqnarray}}

\newcommand{\bm}[1]{{{\rm\bf #1}}}
\begin{document}
\vspace*{4cm}
\title{ITERATIVE MAP-MAKING FOR SCANNING EXPERIMENTS}

\author{ S. Prunet$^1$, C. B. Netterfield$^2$, E. Hivon$^3$, B. P. Crill$^3$ }

\address{${}^1$ CITA, Mc Lennan Labs, University of Toronto \\
  ${}^2$ Dept. of Astronomy, Mc Lennan Labs, University of Toronto \\
  ${}^3$ Caltech Observational Cosmology, Pasadena \\}

\maketitle
\abstracts{
We describe here an iterative method for jointly estimating the 
noise power spectrum from a scanning experiment's time-ordered
data, together with the maximum-likelihood map. 
We test the robustness of this method on simulated 
datasets with colored noise, like those of bolometer receivers
in CMB experiments. }

\section{Introduction}

The map-making problem for CMB anisotropy measurements was first considered
in the context of the COBE-DMR mission \cite{Lin94,Wri94}. 
It was further extended to quick algorithms for differential measurements \cite{Wri96a} 
taking into account $1/f$ noise \cite{Wri96b}.
Given the size of the upcoming datasets ($\geq 10^6$ pixels for {\em MAP}) it is
essential that the map-making algorithm remains quick (typically $O(n_d\log L_N)$
where $n_d$ is the number of time-samples, and $L_N$ the effective noise-filter
length). Another related question, which has been pioneered by \cite{Fer99}, is
how to determine the noise statistical properties from the data itself.
We present a map-making method based on Wright's fast algorithm to compute
iteratively a minimum variance map together with an estimate of the detector
noise power spectrum which is needed to compute unbiased power spectrum estimators
of the cosmological signal. 

\section{Iterative mapmaking - Application to simulations}
\subsection{Method}

Following Tegmark \cite{Teg97}, we model the data stream in the following way:
\begin{equation}
\bm{d_t} = \bm{P_{tp}\Delta_p} + \bm{n_t}
\end{equation}
where $\bm{\Delta_p}$ is a pixelized version of the {\em observed} sky (i.e. convolved
by the experimental beam), and $\bm{n_t}$ is the detector noise after primary
deconvolution of any filter present in the instrumental
chain (e.g. bolometer time constant, read-out filters). 
We will assume here that the experiment is a total power measurement,
i.e. that the pointing matrix $\bm{P_{tp}}$ contains only one non-zero element
per row. 
We now want to estimate the minimum variance map from this data, ie the map
that minimizes $\chi^2 = ({\bm d}-{\bm P\Delta})^\dagger{\bm N}^{-1}({\bm d}-{\bm P\Delta})$.
The solution is given by:
\begin{equation}
\label{eq1}
\tilde\Delta = \left({\bm P}^\dagger{\bm N}^{-1}{\bm P}\right)^{-1}
{\bm P}^\dagger{\bm N}^{-1}{\bm d}
\end{equation}
A few remarks are necessary at this point. First, the matrix to be inverted
is huge ($n_{pix}\times n_{pix}$) so that an iterative linear solver is needed.
Secondly, the noise correlation matrix $N^{-1}_{tt'}$ has to be determined from
the data itself. To make this tractable we assume that, at least over subsets of
the time-stream, the noise is reasonably stationary, so that the multiplication by
$\bm{N}^{-1}$ becomes a convolution operator, in other words that it is diagonal
in Fourier space \footnote{This is actually only approximately true since a convolution
operator is a circulant matrix, ie it assumes that the time-stream has periodic boundary
conditions; this is however a rather good approximation for a time-stream much longer
than the effective length of the noise-filter \cite{Teg97}}.
We thus implemented the following algorithm:\\
\begin{center}
\fbox{\hskip 1cm\parbox[c]{8cm}{
{\bf for each stationary noise subset}
\begin{itemize}
\item{$ \bm{n}^{(j)} = \bm{d} - \bm{P\tilde\Delta}^{(j)} \Rightarrow \bm{N}^{(j)-1} = 
\langle\bm{nn}^\dagger\rangle^{-1} $}
\item{$ \bm{\tilde\Delta}^{(j+1)} - \bm{\tilde\Delta}^{(j)} = 
\left(\bm{P^\dagger W^* P}\right)^{-1}\bm{PN}^{(j)-1}\bm{n}^{(j)}$}
\end{itemize}
{\bf endfor}}}
\end{center}
\vskip 0.5cm
To keep the algorithm as fast as possible, we took $\bm{W^*}$ to be diagonal 
and constant, so that $\bm{P^\dagger W^* P}$ is diagonal, with each element
beeing equal to the number of observations per pixel, up to a multiplicative
constant. The choice of the numerical value of this constant is only important
for the convergence properties of the algorithm. 
In this form, the algorithm is very similar to a Jacobi iterative solver
(in which we would have $\left(diag\{\bm{P^\dagger N}^{-1}\bm{P}\}\right)^{-1} $
instead of $\left(\bm{P^\dagger W^* P}\right)^{-1}$). It has the additional advantage
of being very easy to compute, for very similar convergence properties.
Since we made the assumption that each noise matrix is diagonal in Fourier space,
all time-domain operations are done in Fourier space using FFTs, thus reducing
the number of floating point operations to $O(L_N\log L_N)$ for each subset.

The advantages of this iterative method are obvious: it is fast ($O(n_d\log L_N)$
operations) and cheap in memory ($O(L_N)$ storage). However, since we
try to estimate both the noise power spectrum and the map in a leap-frog manner,
the convergence properties must be studied by means of numerical simulations.

\subsection{Simulations}

To test the convergence properties of the algorithm, we simulated a scanning
experiment on a fake sCDM map with a gaussian beam of $10.5$ arcmin FWHM, and noise
whose power spectrum is $P(f) \propto (1+f_0/f)^2$. To simulate the contamination 
of low frequencies by possible scan-synchronous systematics, we high-pass filtered
the generated time-stream with a step filter. This has the effect of formally replacing
the pointing matrix $\bm{P}$ by $\bm{FP}$ where $F$ is the step filter; however,
looking at Eq.~\ref{eq1}, this is completely equivalent to leaving $\bm{P}$ unchanged,
and replacing $\bm{N}^{-1}$ by $\bm{F^\dagger N}^{-1}\bm{F}$, which is straightforward
in Fourier space.

To be more specific, we used a spherical cap of $\sim 4\times 10^4$ HEALPix pixels 
of $7'$ size \cite{Gor98} generated from
an sCDM COBE-normalized power spectrum, to generate a $\sim 6\times 10^6$ sample
time stream. The sampling rate used is $60$ Hz, and the scan velocity
$1$ deg/s. The sky coverage of the simulation is shown in Fig.~$1$.
The same figure shows the input map used for the simulation, as well as the error
map from the naive (coadded) map-making and the iterative map-making after $10$ iterations.
One can see from those that the striping is much reduced in the iterative case,
the only remaining feature in the error map being a large-scale mode that was lost
in the high-pass filtering of the time-stream. Fig.~$2$ shows the (inverse) noise 
power spectrum estimate compared to the data power spectrum and the true power spectrum.
We can see that the estimated PS is very close to the input one.

As in any iterative linear solver, the convergence of the solution, decomposed
on the (noise-) matrix eigenmodes, is a function of the associated eigenvalue.
Thus the noisiest pixel modes (usually at large scales since the time-stream
is high-pass filtered, either by hand or as part of the algorithm if the noise
is higher at low frequency) take (exponantially) more time to converge.
The problem scales as the noise matrix condition number, which is a direct function of
the noise power spectrum dynamical range and of the scanning strategy.
We propose to accelerate the convergence of the algorithm with the implementation
of a multi-grid like method where the time-stream, as well as the pixels, get rebinned.
This method will be described in a future paper.

\begin{figure}[ht]
\label{fig1}
\vbox{
 \vskip -2cm 
 \hbox{
  \psfig{width=0.45\textwidth,file=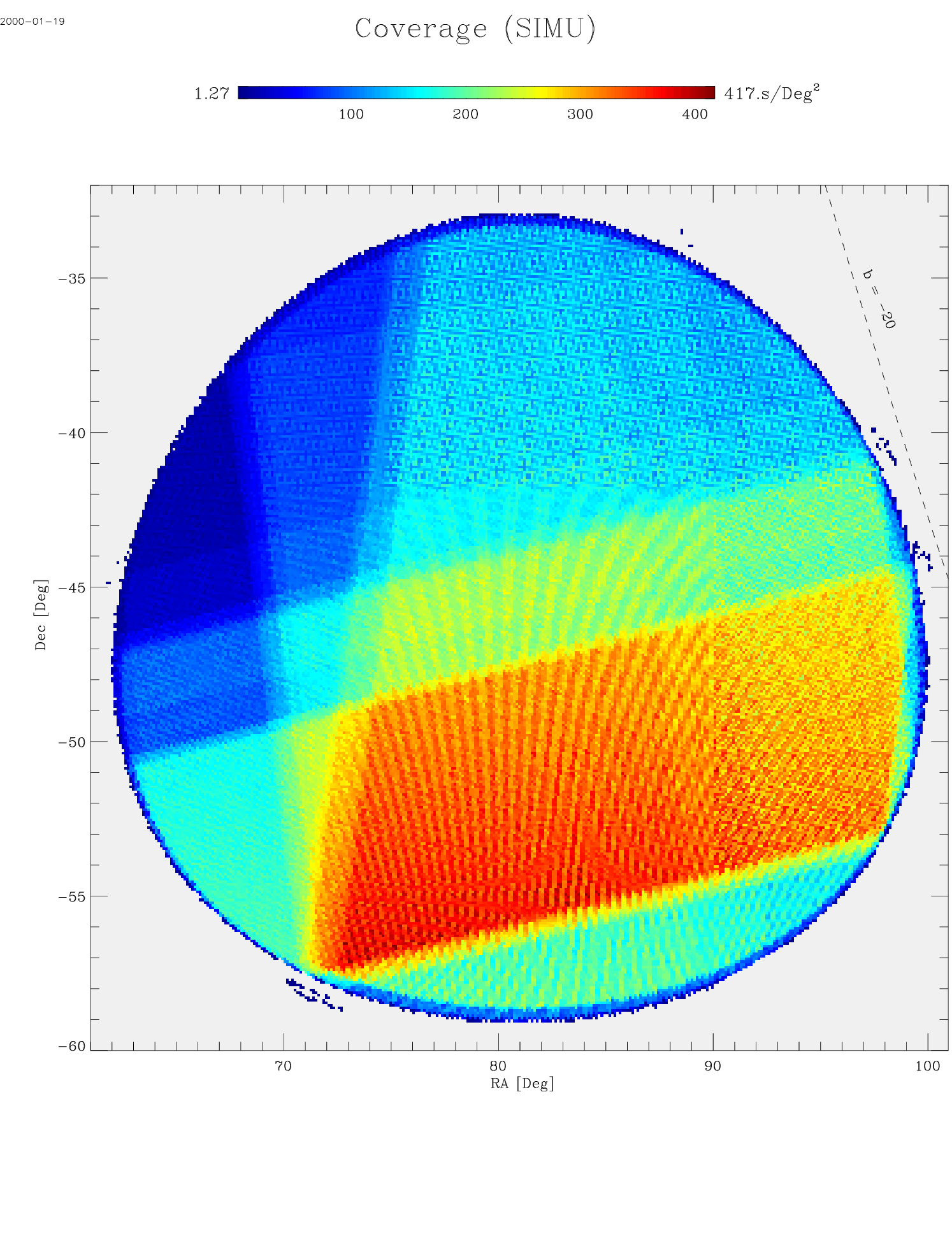}
  \psfig{width=0.45\textwidth,file=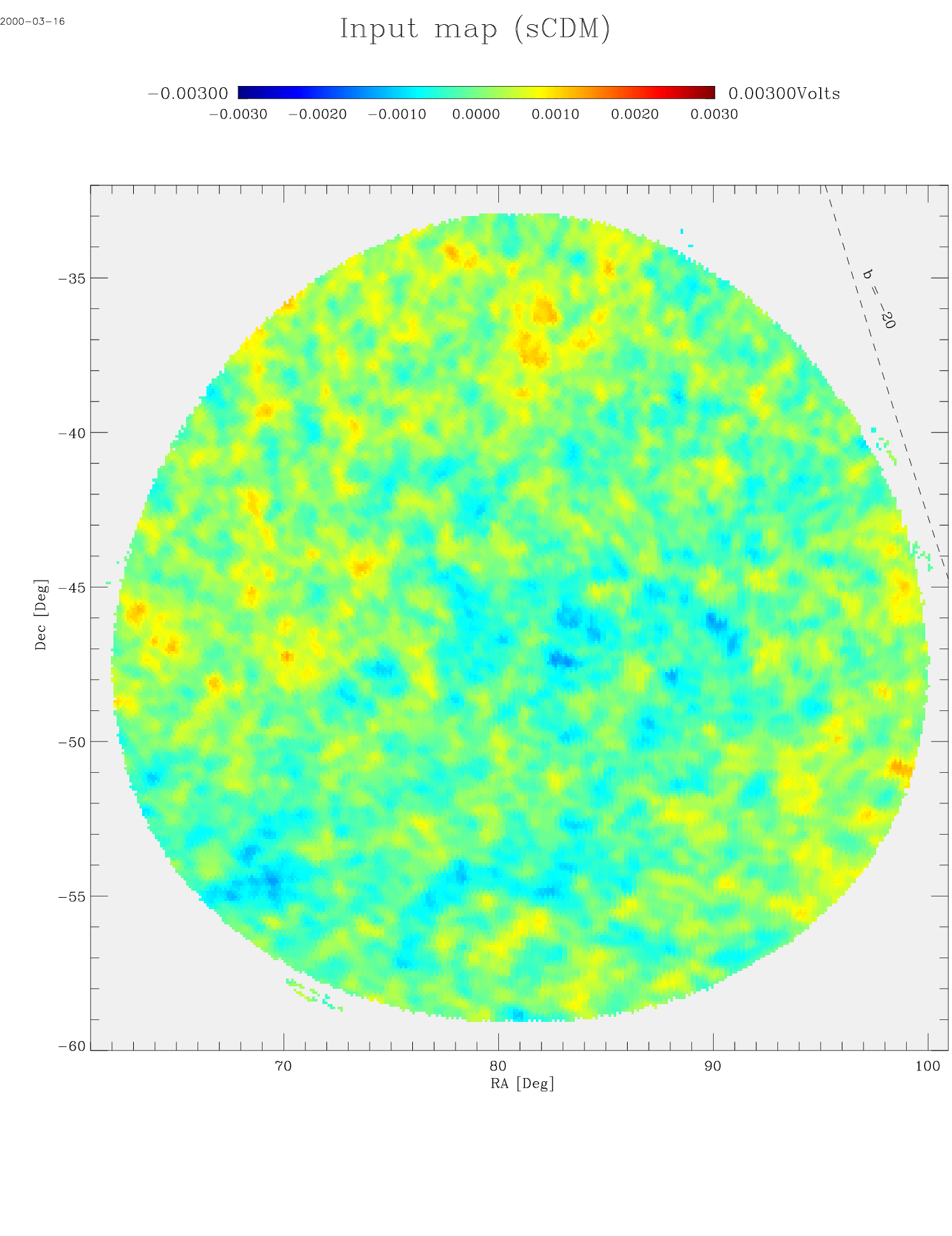}
 }
 \vskip -0.5cm
 \hbox{  
  \psfig{width=0.45\textwidth,file=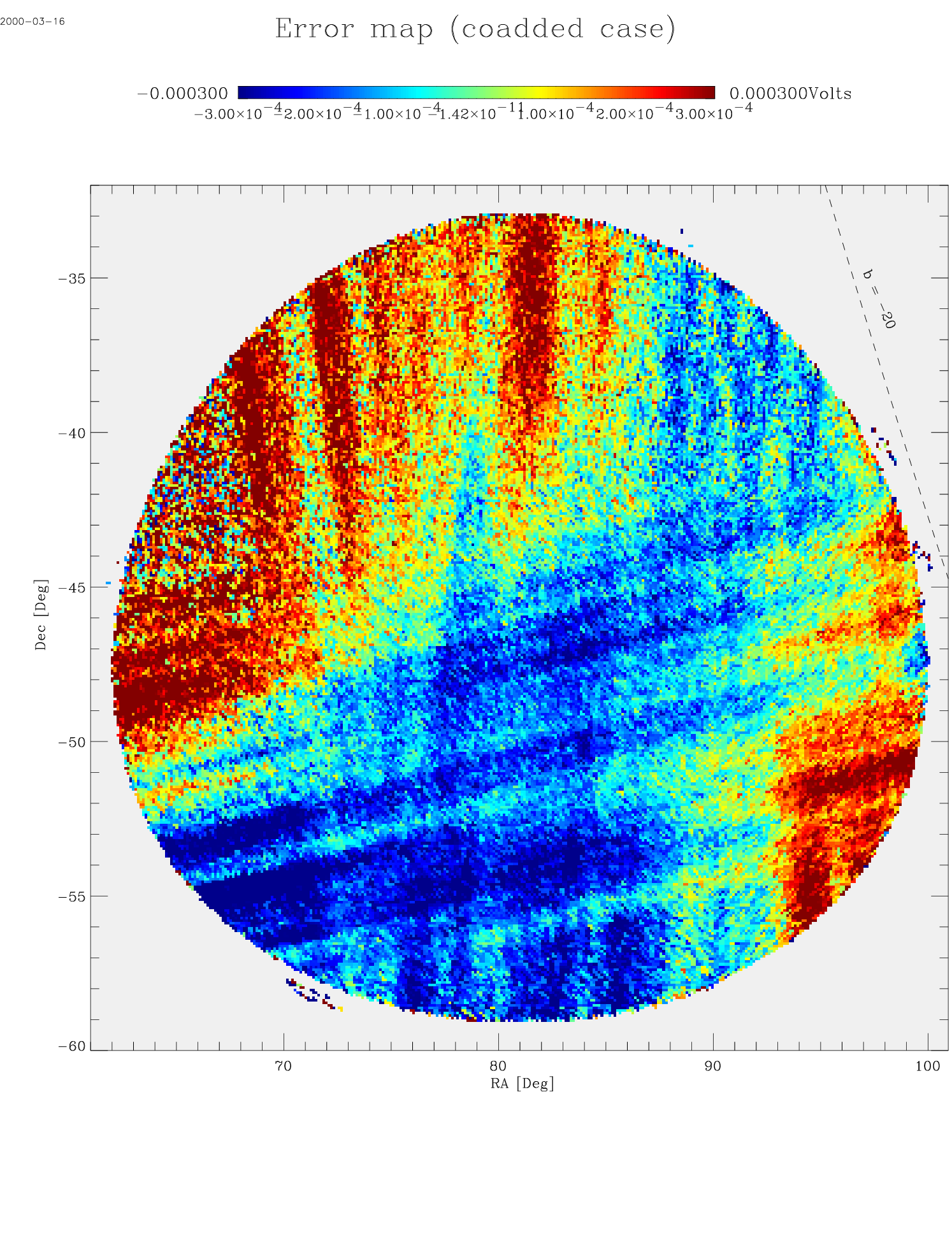}
  \psfig{width=0.45\textwidth,file=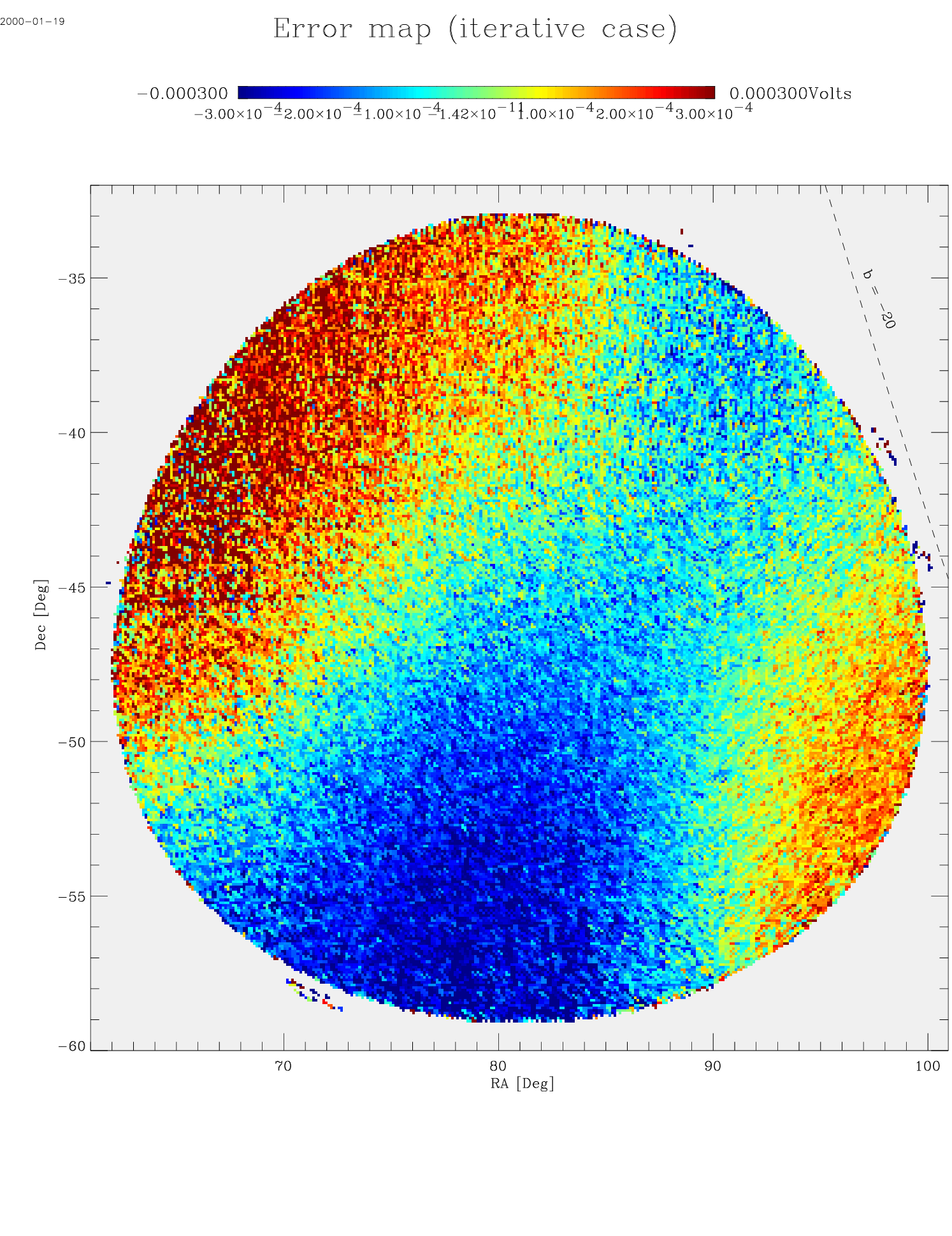}
 }
\vskip -1cm
\caption{\small Simulations- Upper left: Coverage - Upper right: input map - 
Lower left: error map (coadded case) - Lower right: error map (iterative case).}
}
\end{figure}
\begin{figure}[h]
\label{fig2}
\vskip -0.5cm
\centering\psfig{width=0.5\textwidth,file=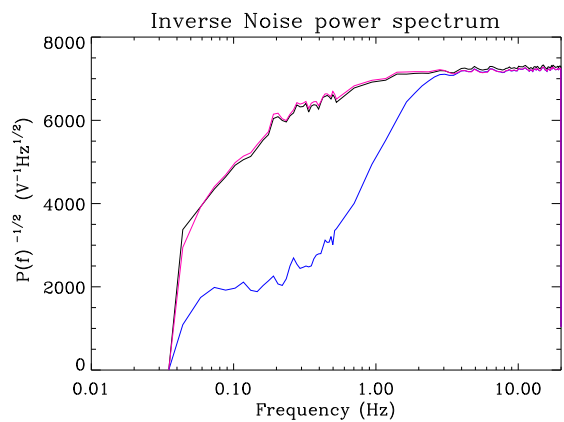}
\vskip -0.5cm
\caption{\small Inverse power spectra of noise in the time-stream. Black is the input noise,
blue is the data, and red is the estimation.}
\end{figure}

\section{Conclusions - Perspectives}

We described a fast iterative map-making method
to simultaneously generate the maximum-likelihood map and the noise power spectrum
from a scanning experiment time stream. We tested its convergence properties 
on simulations, and concluded that, except for the spatially largest 
(and ill conditioned) modes, the map and noise power spectra converge
very quickly. We propose to cure
the convergence of this large scale modes by applying a multi-grid method 
to be described in a future publication.

\end{document}